\documentclass[12pt]{iopart}
\usepackage{iopams}
\usepackage{graphicx,cite,color,epsf}

\begin{document}
%----------------------------------------------------------------------------
%                              TITLE
%----------------------------------------------------------------------------
\title[Multibranched polymers]{Toy models of multibranched polymers: opened vs. circular structures}

\author{K. Haydukivska$^{1,2}$ and V. Blavatska$^1$}
\address{$^1$Institute for Condensed Matter Physics of the National Academy of Sciences of Ukraine 79011 Lviv, Ukraine}
\address{$^2$
Institute of Physics, University of Silesia, 75 Pu\l{}ku Piechoty 1, 41-500 Chorz{\'o}w, Poland}
\ead{blavatskav@gmail.com}

\begin{abstract} 
We study the conformational   properties of complex Gaussian polymers containing  $f_c$ linear branches and $f_r$ closed loops, periodically tethered at $n$ branching points to either a linear polymer backbone (generalized bottlebrush structures) or closed polymer ring  (decorated ring structure).
%We apply to different methods to study the %influence of the branching parameters onto 
 %the universal properties of he structures. %We are interested in the properties of %dilute solutions in regime of %$\theta$-solvent. 
Applying the path integration method, based on Edwards continuous chain model, we obtain in particular the exact values for the size ratios comparing the gyration  radii of considered complex structures and linear chains of the same total molecular weight, as functions of $n$, $f_c$ and $f_r$. Compactification of the overall effective size of branched macromolecules with the increasing number of loops  is quantitatively confirmed. Our results are supported by numerical estimates obtained by application of Wei's method.
%for which the values of shape %characteristics were also obtained.   
    \end{abstract}
\submitto{Journal of Physics A: Mathematical and Theoretical}

\date{\today}
\maketitle
\section{Introduction}

Bottlebrush polymers are  important representatives of the class of
 complex multibranched  structures, where the sets of $f_c$ side chains are periodically tethered at $n$ branching points to a linear polymer backbone \cite{Sheiko08}. The  properties of such molecules are  governed by competition of the steric repulsion between the side chains and the configurational entropy of the main backbone chain. In particular, this leads to enhanced stiffness of the backbone  which causes the considerable elongation of molecule in solvent 

 %as compared to individual linear chain of the same molecular %weight 
 \cite{Tereo99,Kawaguchi98}; the side-chain molecules attend  the extended cylindrical shapes in solvents \cite{Liang17,Lopez15}.
 In turn, the molecular entanglement in melts of such molecules is significantly suppressed and their conformational  degrees of freedom  are lowered \cite{Namba00,Sheiko08,Rzayev12}, which dictates the viscoelastic properties of such molecules. 
 %This reduced entanglement effect
%is exploited e.g. in designing low modulus elastomers capable of strain-hardening and %sustaining large deformation %\cite{Cai15,Liang18}. 
%BBPs are also characterized by 
%unique self-assembly behaviors in solution and bulk states; and
%have been observed to self-assemble into spherical \cite{Hernandez},
%rod-like \cite{Pang}, lamellar \cite{Shi} %and other structures.
 These peculiarities can be processed and controlled in a wide range  by varying the structural parameters such as grafting density of the side chains,  their lengths and position along the backbone.

\begin{figure}[b!]
\begin{center}
\includegraphics[width=63mm]{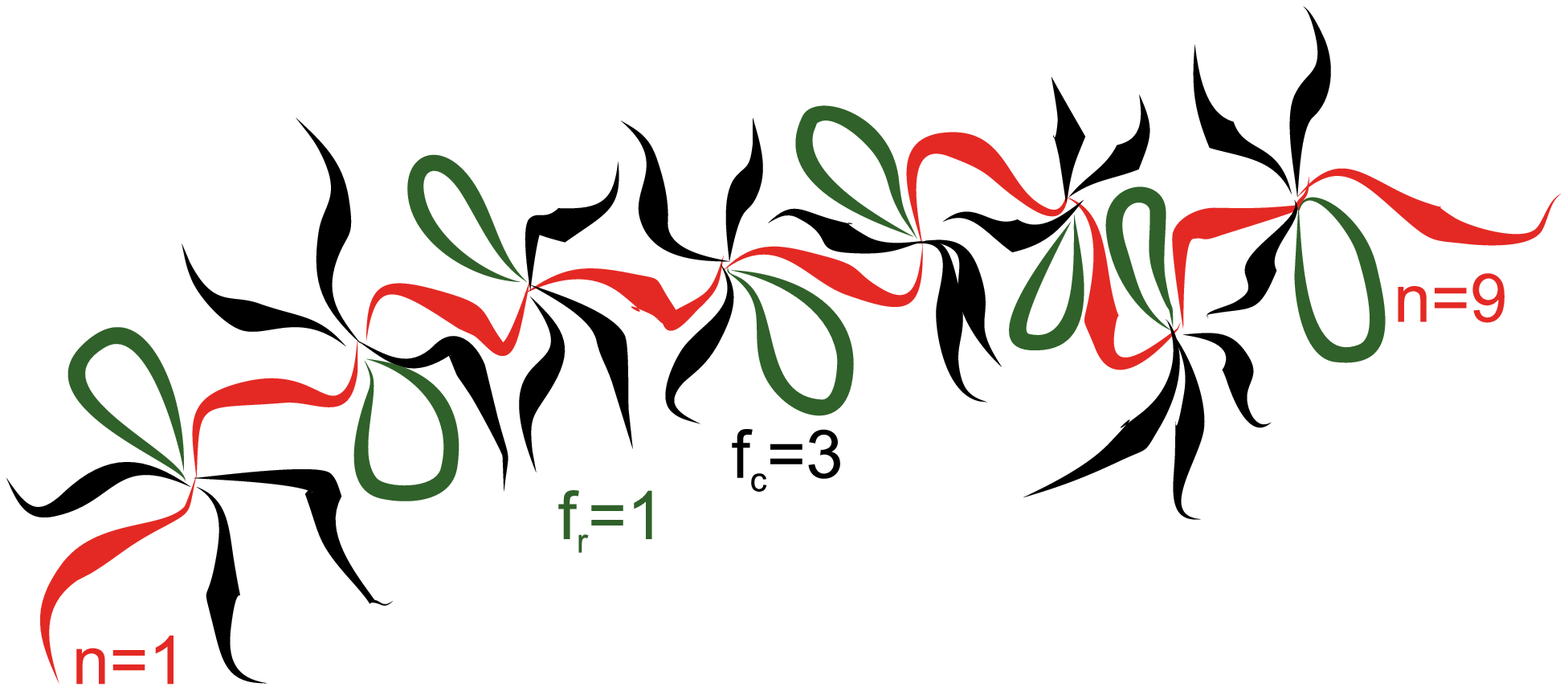}
\includegraphics[width=53mm]{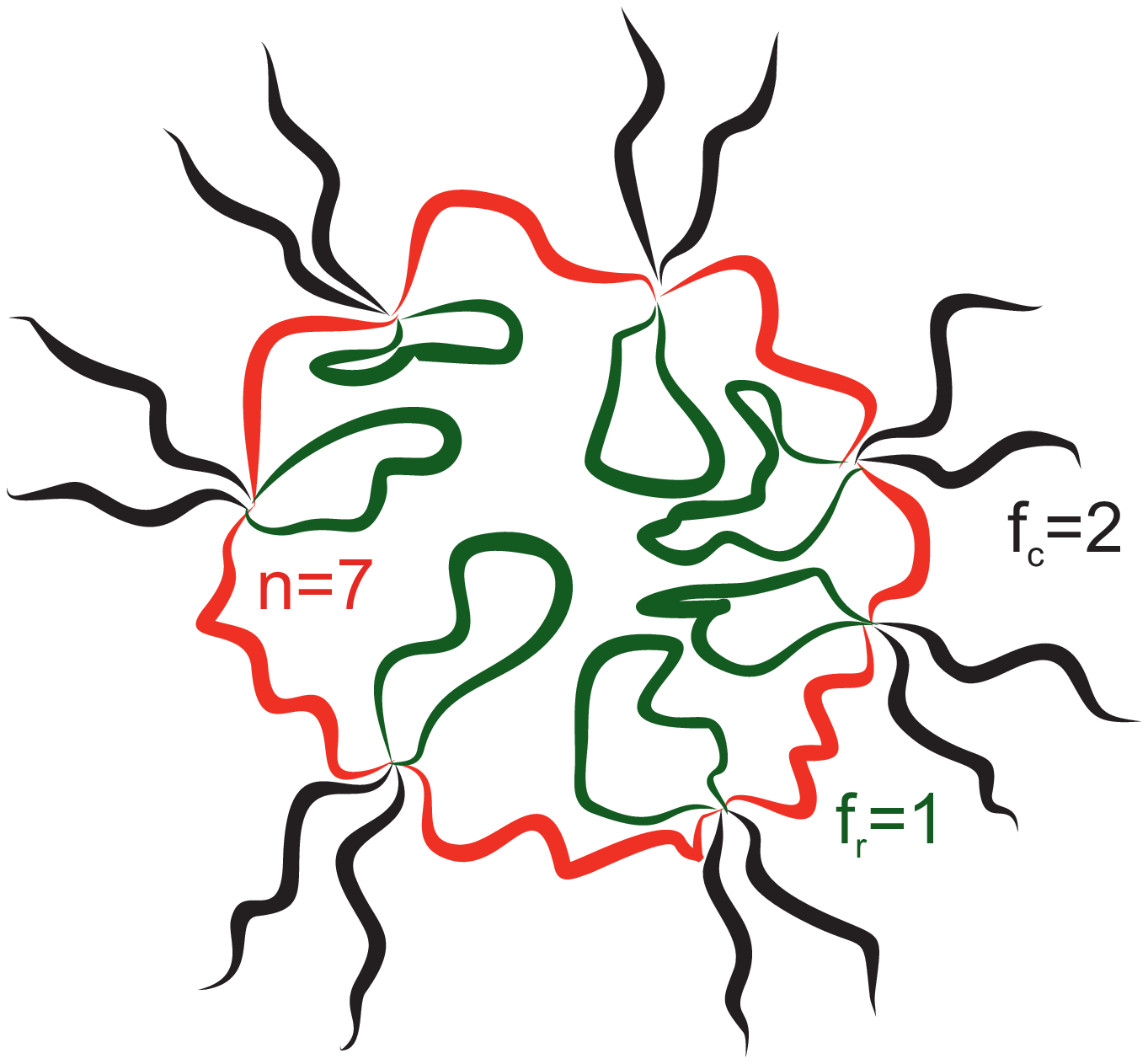}
\caption{ \label{Brush} Schematic representation of the open (left) and circular (right) bottlebrush polymer containing $f_c$ linear side chains and $f_r$ side chains in form of closed loops.  }
\end{center}
\end{figure}

Loop formation in macromolecules plays an important role in a number of biochemical
processes such as stabilisation of globular proteins \cite{Pace88,Nagi97}, DNA  compactification in the nucleus \cite{Fraser06,Simonis06,Dorier09} etc. Cyclic (or ring) macromolecules in general exhibit a number of unique properties  in comparison to their linear analogues. In particular, they are characterized by a more
compact coil conformation due to their lower conformational degree of freedom and thus diffuse much faster in melts than linear polymer chains of the same molecular weight, thus exhibiting very different viscoelastic and phase transition behavior \cite{r1,r2}. Cyclic polymers are thus finding their application as  viscosity modifiers, since their addition in small amounts to linear matrices 
increases the viscosity \cite{Vlas,Kenna}.
%In technologies, cyclic polymers can be %used as templates for the assembly of
%nanoparticles \cite{Yu}, the surface %modifiers to prevent nanoparticle
%aggregation \cite{Morgese,Ver}.
Numerous analytical and numerical studies have been conducted
to analyse the cyclisation probability and loop size distributions in long flexible
macromolecules \cite{Chan89,Rubio93,Redner80,Hsu04,Duplantier89, Wittkop96,Zhou01,Toan06,Shin14}. In particular, the  so-called rosette polymers, containing a single branching point with $f_c$ linear chains and $f_r$ circular loops eliminating have been studied recently in our previous works \cite{Metzler,Paturej20}. In this concern, it is worthwhile to consider the
generalised multibranched  structures, where some amount $f_r$ of side chains at each the branching point are closed and form loops (see Fig. \ref {Brush}).
One more closely related architecture of a branched polymer  is a  so-called decorated ring polymer: a bottlebrush polymer that has its backbone looped (see Fig. \ref {Brush}). The 
cyclic polystyrenes with multiple  branching have been synthesized and analyzed recently in Refs. \cite{Hossain1,Hossain2}.  

In our present study, we will focus on conformational properties of the above mentioned complex polymer  structures in the regime of dilute solution. In particular, the size ratio of gyration radii of a branched polymer  $\langle R_g^2 \rangle_{\rm branched}$ to that
of the linear chain $\langle R_g^2 \rangle_{\rm chain}$ of the the same total molecular weight
\begin{equation}
g=\frac{\langle R_g^2 \rangle_{\rm branched}}{\langle R_g^2 \rangle_{\rm chain}} \label{gratio}
\end{equation}
is often used to characterize the impact of the complex topology of structure on the effective elongation  in solvent (it is also referred to as the shrinking factor). Here and below, $\langle (\ldots) \rangle$ denotes averaging over an ensemble of possible polymer conformations.
Originally, this value was introduced by Zimm \cite{Zimm49} and evaluated for the so-called star polymer structures, containing only one branching point, as well as for a single  closed ring.
Another important quantity that characterizes  the effective size of a polymer coil is hydrodynamic radius $R_h$, which is directly obtained in dynamic light scattering experiments.
To compare $R^2_g$ and $R^{-1}_h$, it is convenient to introduce the universal size ratio
\begin{equation}
\rho=\sqrt{\langle R_g^2 \rangle} /\langle R_H \rangle. \label{hratio}
\end{equation}
 The exact analytical result for the ratio (\ref{hratio})
in $d=3$ dimensions have been evaluated for the linear chain \cite{Zimm49,burchard,dunweg} and single closed ring \cite{burchard,fukatsu,Uehara2016}. 
Note,  that the above mentioned quantities can be estimated exactly in the case of ideal Gaussian polymers without taking into account an excluded volume effect.
Recently, both the analytical and numerical  estimates for the universal ratios (\ref{gratio}), (\ref{hratio}) have been obtained in Refs. \cite{Metzler,Paturej20,Budkov}  
for so-called rosette structure, containing single branching point with $f_r$ closed rings and $f_c$ linear chains, which  can be considered as bottlebrush polymers with $n=2$. { Note that the probability distribution functions of the gyration radii of Gaussian star and rosette structures have been derived analytically in Ref. \cite{Budkov} as well. } 
 In this approach, an estimate for the Gaussian 
bottlebrush polymers containing $n$ branching points of functionality $f$ is found in Ref. \cite{Nakamura}. {There is a  number of  studies dedicated to conformational properties (and in particular the size ratios) of bottlebrush polymers with linear side chains  \cite{Ma94,ZHIPING92,Paturej16,Chremos18,Pan22,Theodorakis}. In particular, a thorough comparative study of size properties of bottlebrush polymer structures with linear, star and ring polymers is performed  in Ref. \cite{Chremos18}.  The aim of the present study is to continue these investigations to the case when the side chains form the closed loops. This problem is closely related with the process called loop extrusion in chromatin \cite{Alipour12,Fudenberg16,Goloborodko16,Mirny21}, serving as one of the mechanisms of compaction of long DNA molecules in nuclei. Moreover, we will consider the case when the backbone chain form a closed loop  (the case of ``decorated ring''), which is of great interest in particular since DNA often has a closed ring shape \cite{DNA1,DNA2}.} 

%Also size ratios introduced above can qualitatively describe the change in shape of the macromolecule.  For a more quantitative description it is traditional to consider so called universal shape characteristics that are invariants of gyration tensor \cite{Aronovitz86,Rudnick86}. The aspherisity $\langle A \rangle$ being one of them. It is always positive and has the maximum possible value of $1$ for a absolutely rigid chain. In the case of the simple linear chain in Gaussian case it was approximated to be $\langle A \rangle_{chain}=0.377$ for the chain and $\langle A \rangle_{chain}=0.261$ for the ring \cite{Bishop88}. 

The layout of the paper is as follows. In Section \ref{M} we introduce the continuous chain model for both structures that is followed by analytical calculation in Section \ref{Calc}. The Wei's method is introduced in Section \ref{met1} that is followed by Results and Discussions is Section \ref{RD}. We finish this work with concluding remarks in Section \ref{C}. 

\section{The model}
\label{M}
An analytical description is performed in the framework of continuous chain model. Here, the individual polymer chain is presented as a trajectory of length $L$ parameterised by a radius vector $\vec{r}(s)$ with $s$ changing from $0$ to $L$. The trajectories connectivity is determined by the hamiltonian\cite{desCloiseaux}:
\begin{equation}
H^0=\frac{1}{2}\int_0^L \left(\frac{d\vec{r}(s)}{ds}\right)^2. \label{H_0}
\end{equation}

We consider a generalized case when each of the $n-1$ branching points on the backbone linear chain can contain $f_c$ side chains and $f_r$ rings. Mathematically this topology is defined by the partition function:
\begin{eqnarray}
&&Z_{bb}=\frac{1}{Z_0}\int D\vec{r} \prod_{k=1}^{n-1}\delta(\vec{r_k}(L_b)-\vec{r_{k+1}}(0))\nonumber\\
&&\times\prod_{i=1}^{f_c+f_r}\delta(\vec{r_k}(L)-\vec{r_i}(0))\prod_{j=1}^{f_r}\delta(\vec{r_j}(L)-\vec{r_j}(0)){\rm e}^{-\sum_{j=1}^{F}H_j^0}. \label{Zbb}
\end{eqnarray}
Here, the first set of $\delta$-functions describes the backbone that consists of $n$ segments with each consecutive segment starting at the end of the previous one. The second and the third sets of $\delta$-functions stands for the fact that each of the branching points serves as a starting point for $f_c+f_r$ outgoing trajectories. Note that here we consider the case when all the side chains are of the same length  $L$, however the segments of the backbone are of the different length $L_b$ such that $\lim_{L\rightarrow\infty}(L_b/L)=a$. In this case the total length of the backbone is $nL_b$. The total number of trajectories for this structure being $F=n+(n-1)(f_c+f_r)$.

For the case of decorated ring, that is a bottlebrush polymer that has two end points of its backbone chain looped, the partition function is presented as:
\begin{eqnarray}
&&Z_{dr}=\frac{1}{Z_0}\int D\vec{r}\,\, \delta(\vec{r_n}(L_b)-\vec{r_{0}}(0))\nonumber\\ &&\times\prod_{k=0}^{n-1}\delta(\vec{r_k}(L_b)-\vec{r_{k+1}}(0))\prod_{i=1}^{f_c+f_r}\delta(\vec{r_k}(L)-\vec{r_i}(0))\nonumber\\
&&\times\prod_{j=1}^{f_r}\delta(\vec{r_j}(L)-\vec{r_j}(0)){\rm e}^{-\sum_{j=1}^FH_j^0}. \label{Zdr}
\end{eqnarray}
Here, an  additional condition  $\delta(\vec{r_n}(L)-\vec{r_{0}}(0))$ stands for its end points  looping. The total number of trajectories in this case is $F=n(f_c+f_r+1)$.

Within this model, any observable of interest can be calculated as an average over an ensemble of all possible trajectories:
\begin{equation}
\langle(\ldots)\rangle = \frac{1}{Z}\int D\vec{r} {\rm e }^{-\sum_{j=1}^{F}H_j^0}, \label{Zaver}
\end{equation}
where  $Z$ is a partition function given either by Eq. (\ref{Zbb}) or Eq. (\ref{Zdr}).

\section{Gaussian approximation for analyzing the size characteristics of polymers}
\label{Calc}

\subsection{Radius of gyration and hydrodynamic radius definitions and calculation strategies}

In terms of continuous chain model introduced above,  the size measures of polymer structure can be defined as:
\begin{eqnarray}
&&\langle R^2_g\rangle = \frac{1}{2(FL)^2}\sum_{i,j=1}^F\int_0^L\int_0^L\langle(\vec{r_i}(s_2)-\vec{r_j}(s_2))^2\rangle,\label{RGdef}\\
&&\langle R^{-1}_h\rangle = \frac{1}{(FL)^2}\sum_{i,j=1}^F\int_0^L\int_0^L\langle|\vec{r_i}(s_2)-\vec{r_j}(s_2)|^{-1}\rangle,\label{RHdef}
\end{eqnarray}
here $s_2$ and $s_1$ stand for positions   along and referred as ``restriction points''.

The calculation of the gyration radius within this model is performed by making  use of identity \cite{desCloiseaux}:
\begin{eqnarray}
&&\langle(\vec{r}_i(s_2)-\vec{r}_j(s_1))^2\rangle = - 2 \frac{d}{d|\vec{k}|^2}\xi(\vec{k})_{\vec{k}=0},\nonumber\\
&&\xi(\vec{k})\equiv\langle{\rm e}^{-\iota\vec{k}(\vec{r}_i(s_2)-\vec{r}_j(s_1))}\rangle.\label{identity_g}
\end{eqnarray}
The crucial point in the calculation of the hydrodynamic radius is exploiting the following equality \cite{Haydukivska14}:
\begin{equation}
|\vec{r}|^{-1} = (2\pi)^{-d}   \int\, d\vec{k}\, 2^{d-1} \pi^{\frac{d-1}{2}} \Gamma\left(\frac{d-1}{2}\right) \, k^{1-d}{\rm e}^{i\vec{r}\vec{k}}.
\end{equation}
With this Fourier transformation the average in the integrand  in expression (\ref{RHdef}) can be rewritten as:
\begin{eqnarray}
&&\langle|\vec{r}_i(s_2)-\vec{r}_j(s_1)|^{-1}\rangle  = (2\pi)^{-d}   \int\, d\vec{k}\, 2^{d-1} \pi^{\frac{d-1}{2}} \Gamma\left(\frac{d-1}{2}\right) \, k^{1-d} \xi(\vec{k}).\label{identity_h}
\end{eqnarray}

Note that in both identities the averaging  is performed for the same expression $\langle{\rm e}^{i\vec{k}(\vec{r}_i(s_2)-\vec{r}_j(s_1))}\rangle$. In the Gaussian case, performing the averaging according to (\ref{Zaver}) for a single polymer chain of length $L$ leads to an expression:
\begin{equation}
\langle{\rm e}^{i\vec{k}(\vec{r}_i(s_2)-\vec{r}_j(s_1))}\rangle={\rm e}^{-\frac{\vec{k}^2}{2}(s_2-s_1)},
\end{equation}
and the application of the identities (\ref{identity_g}) and  (\ref{identity_h}) results in:
\begin{eqnarray}
&&\langle(\vec{r}_i(s_2)-\vec{r}_j(s_1))^2\rangle=d(s_2-s_1),\\
&&\langle|\vec{r}_i(s_2)-\vec{r}_j(s_1)|^{-1}\rangle= \frac{\Gamma\left(\frac{d-1}{2}\right)}{\sqrt{2}\Gamma\left(\frac{d}{2}\right)}(s_2-s_1)^{-1/2}.
\end{eqnarray}
After performing integrating over $s_1$ and $s_2$ and dividing by $L^2$ the final expressions for a single polymer chain are:
\begin{eqnarray}
&&\langle R^2_g\rangle_{{\rm chain}}=\frac{dL}{6},\\
&&\langle R^{-1}_h\rangle_{{\rm chain}}=\frac{\sqrt{2}\Gamma\left(\frac{d-1}{2}\right)}{\Gamma\left(\frac{d}{2}\right)}\frac{4}{3}L^{-\frac{1}{2}}.
\end{eqnarray}
The size ratio between those two size characteristics as defined by expression (\ref{hratio}) is given by:
\begin{equation}
\rho_{{\rm chain}}=\frac{\sqrt{\langle R^2_g\rangle}_{{\rm chain}}}{\langle R^{-1}_h\rangle_{{\rm chain}}^{-1}}=\frac{\sqrt{2d}\Gamma\left(\frac{d-1}{2}\right)}{\Gamma\left(\frac{d}{2}\right)}\frac{4}{3\sqrt{6}},
\end{equation}
which reproduces a well-known value \cite{Teraoka02}. 

\subsection{Calculation of size characteristics}

\begin{figure}[t!]
\begin{center}
\includegraphics[width=100mm]{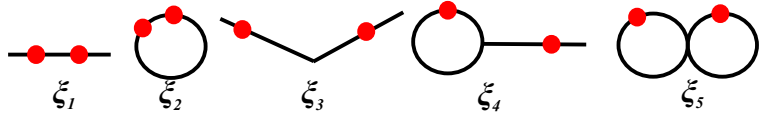}
\caption{ \label{D:1} Schematic representation of the contributions into the gyration and hy\-drodynamic radius of rosette polymer. Here, solid line represents polymer trajectory and bullets represent restriction points.}
\end{center}
\end{figure}

\begin{figure}[b!]
\begin{center}
\includegraphics[width=100mm]{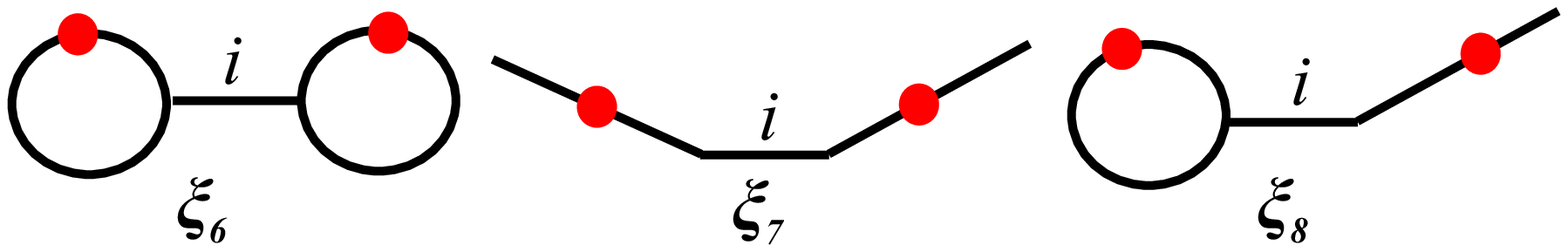}
\caption{ \label{D:2} Schematic representation of the contributions into the gyration and hydrodynamic radius of bottlebrush polymer. Notations are the same like on Fig. \ref{D:1}.}
\end{center}
\end{figure}

It is convenient to use a diagram technique for schematic presentations of contribution into $\xi(\vec{k})$. 
In the case of  bottlebrush structure,  the set of diagrams consists of five diagrams given on Fig. \ref{D:1} (those corresponding to rosette structures and being  calculated previously in our works \cite{Metzler,Paturej20}) and three additional ones presented on Fig. \ref{D:2}. Here,  the length of the each segment plays an important role and has to be accounted for. The analytical expressions corresponding to all the diagrams  are provided in the Appendix.

 The size characteristics of interest can in general be presented as:
\begin{eqnarray}
&& R_{x}= \frac{L^p}{(an+(f_r+f_c)(n-1))^2}\left((n-1)\left[f_c\xi_{1,x}(1)+f_r\xi_{2,x}(1)\right.\right.\nonumber\\&&\left.\left.+\frac{f_c(f_c-1)}{2}\xi_{3,x}(1,1)+\frac{f_r(f_r-1)}{2}\xi_{5,x}(1,1)+f_rf_c\xi_{4,x}(1,1)\right.\right.\nonumber\\
&&\left.+2f_c\xi_{3,x}(1,a)+\xi_{3,x}(a,a)+2f_r\xi_{4,x}(1,a)\right]+n\xi_{1,x}(a)\nonumber\\
&&+\left.\sum_{i=1}^{n-2}(n-i-1)\left[f_c^2\xi_{7,x}(1,1)+2f_c\xi_{7,x}(1,a)+\xi_{7,x}(a,a)+2f_cf_r\xi_{8,x}(1,1)\right.\right.\nonumber\\
&&\left.\left.+2f_r\xi_{8,x}(1,a)+f_r^2\xi_{6,x}(1,1)\right]\right).\label{gen_b}
\end{eqnarray}
Here, $p$ is equal to $1$ for the gyration radius and $-1/2$ for the hydrodynamic radius and $x$ stands to the indexes $g$ or $h$. In the case of gyration radius the summations are easily performed and the following expression results:
\begin{eqnarray}
&&\langle R^2_g\rangle_{bb}= \frac{dL}{(an+(f_r+f_c)(n-1))^2}\left(\frac{a^3n^3}{6}\right.\nonumber\\
&&+\left.\frac{n-1}{6}\left((an^2-2an+3n-3)f_c^2+(2a^2n^2-a^2n+3an-2)f_c\right.\right.\nonumber\\
&&\left.\left.+\frac{1}{2}(4a^2n^2-2a^2n+2an-1)f_r+(an^2-2an+n-1)f_r^2\right.\right.\nonumber\\
&&\left.\left.+2(an^2-2an+2n-2)f_cf_r\right)\right).\label{Rgbb}
\end{eqnarray}
 With $f_r=0$, the above expression restores that for the bottlebrush polymer with linear side chains, derived in Ref. \cite{Nakamura}.
To evaluate an estimate for the size ratio (\ref{gratio}), we  divide this expression by the value of gyration radius of a single chain with the same total molecular weight, that is    $\langle R^2_g\rangle_{chain}=\frac{dL(an+(f_r+f_c)(n-1))}{6}$. The graphic representation of  resulting expression for some values of $n,f_c,f_r$ are given on Fig. \ref{g_bottle}. 

Correspondingly, we evaluated an expression for hydrodynamic radius:
\begin{eqnarray}
&&\langle R^{-1}_h\rangle_{bb}=\frac{\sqrt{\frac{2}{L}}\Gamma\left(\frac{d-1}{2}\right)}{(f_c+f_r)^3\Gamma\left(\frac{d}{2}\right)}
\left(\frac{4a^{\frac{3}{2}}n}{3}+\frac{n-1}{12}\left((16(2a^{\frac{3}{2}}+f_c^2))(\sqrt{2}-1)\right.\right.\nonumber\\
&&+(32(1+a)^{\frac{3}{2}}-32a^{\frac{3}{2}}-16\sqrt{2})f_c+12\arctan\left(\frac{1}{2\sqrt{a}}\right)(4a+1)f_r\nonumber\\
&&\left.-6f_r\pi(f_r-1)(\sqrt{2}-2)+24\sqrt{a}f_r+3f_rf_c(4-\pi+10\arctan\left(\frac{1}{2}\right))\right)\nonumber\\
&&+\frac{4}{3}\sum_{i=1}^{n-2}(n-i-1)\left(f_c^2(ai+2)^{\frac{3}{2}}+2f_c(ai+a+1)^{\frac{3}{2}}-2f_c(ai+1)^{\frac{3}{2}}(f_c+1)\right.\nonumber\\
&&+(ai+2a)^{\frac{3}{2}}-2(a(i+1))^{\frac{3}{2}}(f_c+1)\left.+(ai)^{\frac{3}{2}}(f_c-1)^2\right)\nonumber\\
&&+2f_cf_r\sum_{i=1}^{n-2}(n-i-1)\left(\frac{1}{2}\arctan\left(\frac{1}{2\sqrt{ai+1}}\right)(4ai+5)\right.\nonumber\\&&\left.-\frac{1}{2}\arctan\left(\frac{1}{2\sqrt{ai}}\right)(4ai+1)\right.\left.+\sqrt{ai+1}-\sqrt{ai}\right)\nonumber\\
&&+2f_r\sum_{i=1}^{n-2}(n-i-1)\left(\frac{1}{2}\arctan\left(\frac{1}{2\sqrt{ai+a}}\right)(4ai+4a+1)\right.\nonumber\\
&&\left.-\frac{1}{2}\arctan\left(\frac{1}{2\sqrt{ai}}\right)(4ai+1)+\sqrt{ai+a}-\sqrt{ai}\right)\nonumber\\
&&-f_r^2\sum_{i=1}^{n-2}\frac{n-i-1}{\sqrt{4ai+2}}\left((4ai+2)\left(\arctan\left(\frac{4ai+1+\sqrt{4ai+2}}{2\sqrt{ai}}\right)\right.\right.\nonumber\\&&\left.\left.+\arctan\left(\frac{\sqrt{4ai+2}-4ai-1}{2\sqrt{ai}}\right)\right)\right.-2\arcsin\left(\frac{1}{\sqrt{4ai+1}}\right)\sqrt{4ai+2}\nonumber\\
&&\left.-2\arctan\left(\frac{1}{2\sqrt{ai}}\right)\sqrt{4ai+2}\right). \label{Rhbb}
\end{eqnarray}
Here $L$ is the length of the side chain and $L_b=aL$.

The ratio (\ref{hratio}) for the bottlebrush structures can be evaluated next on the base of (\ref{Rgbb}) and (\ref{Rhbb}).  The graphic representation of  resulting expression for some values of $n,f_c,f_r$ are presented  on Fig. \ref{ro_b}.

\begin{figure}[t!]
\begin{center}
\includegraphics[width=100mm]{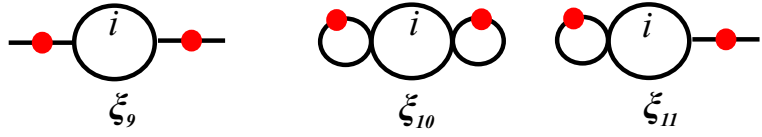}
\caption{ \label{D:3} Schematic representation of contributions into the gyration and hydrodynamic radii of decorated ring polymer. Notations are the same like on Fig. \ref{D:1}.}
\end{center}
\end{figure}

For the case of decorated ring structures,  we again need to consider  three additional diagrams (see fig. \ref{D:3}) besides of those already considered for bottlebrush structures.

\begin{table}[b!]
\begin{tabular}{| c | c | c | c | c |}
   \hline
$f_c$ &  $f_r$ & $n$ & $\rho$\\ \hline
$0$ &  $1$ & $2$ &  $\frac{\sqrt{22}}{32\sqrt{\pi}}\left(3\pi\sqrt{2}-2\pi\sqrt{6}-4\arctan(2\sqrt{2})+2\arcsin\left(\frac{\sqrt{3}}{3}\right)\right.$\\ 
&   &  &  $\left.+9\pi-6\arctan\left(\frac{5}{\sqrt{2}}\right)\right)$\\ \hline
$0$ &  $2$ & $2$ &  $\frac{\sqrt{6}}{36\sqrt{\pi}}\left(3\pi\sqrt{2}-4\pi\sqrt{6}+30\pi-16\arctan(2\sqrt{2})\right.$\\ 
&   &  &  $\left.-24\arctan\left(\frac{5}{\sqrt{2}}\right)+8\arcsin\left(\frac{\sqrt{3}}{3}\right)\right)$ \\ \hline
$0$ &  $1$ & $3$ &  $\frac{\sqrt{69}}{216\sqrt{\pi}}\left(21\pi-3\pi\sqrt{3}-2\arctan\left(\frac{24\sqrt{7}}{103}\right)\sqrt{42}+12\arctan\left(\frac{\sqrt{6}}{4}\right)\right.$\\ 
&   &  &  $\left.+12\arcsin\left(\frac{\sqrt{3}}{\sqrt{11}}\right)\right)$ \\ \hline
$0$ &  $2$ & $3$ &  $\frac{\sqrt{74/\pi}}{486}\left(3\pi\left(18-2^{\frac{3}{2}}-3^{\frac{3}{2}}\right)-8\arctan\left(\frac{24\sqrt{7}}{103}\right)\sqrt{42}\right.$\\ 
&   &  &  $\left.+48\left(\arctan\left(\frac{\sqrt{6}}{4}\right)+\arcsin\left(\frac{\sqrt{3}}{\sqrt{11}}\right)\right)\right)$\\ \hline
$1$ &  $0$ & $2$ & $\frac{\sqrt{35}}{96\sqrt{\pi}}\left(5\sqrt{10}+\sqrt{2}+20-6\sqrt{6}+18\sqrt{2}\arcsin\left(\frac{\sqrt{3}}{3}\right)\right)$\\ \hline
\end{tabular}
\caption{Size ratio  (\ref{hratio}) for a set  of decorated rings with fixed parameters $n$, $f_r$, $f_c$, assuming $a=1$.}\label{Table2}
\end{table}

Similarly to Eq.(\ref{gen_b}), the general expression for size characteristics reads:
\begin{eqnarray}
&& R_{x}= \frac{L^p}{((f_r+f_c+a)n)^2}\left(n\left(f_c\xi_{1,x}(1)+f_r\xi_{2,x}(1)\right.\nonumber\right.\\
&&\left.+\frac{f_c(f_c-1)}{2}\xi_{3,x}(1,1)+\frac{f_r(f_r-1)}{2}\xi_{5,x}(1,1)+f_rf_c\xi_{4,x}(1,1)\right)\nonumber\\
&&+\xi_{1,x}(na)+f_cn\xi_{4,x}(1,an)+f_r n\xi_{5,x}(1,an)\nonumber\\
&&+\left.\sum_{i=1}^{n-1}(n-i-1)\left(f_c^2\xi_{9,x}(1,1)+2f_rf_c\xi_{11,x}(1,1)+f_r^2\xi_{10,x}(1,1)\right)\right).\label{gen_dr}
\end{eqnarray}
For the  gyration radius of decorated ring structure we thus have   
\begin{eqnarray}
&&\langle R^2_g\rangle_{dr}= \frac{dL}{(an+(f_r+f_c)n)^2}\left(\frac{a^3n^3}{12}+\frac{n}{12}\left((an^2-a+6n)f_c^2\right.\right.\nonumber\\
&&\left.\left.+(2a^2n^2+6an-4)f_c+(2a^2n^2+2an-1)f_r\right.\right.\nonumber\\
&&\left.\left.+(an^2-a+2n)f_r^2+(2an^2-2a+8n-8)f_cf_r\right)\right). \label{Rgdr}
\end{eqnarray}
Again, to evaluate the size ratio (\ref{gratio}), we divide this expression by $\langle R^2_g\rangle_{chain}$ of the same total molecular weight, and give some graphical representations on Fig. \ref{g_bottle}.

The expressions for hydrodynamic radius of  decorated ring structure is much more cumbersome and we give here examples of this expression for several simple cases of $n,f_c$ and $f_r$ in Table \ref{Table2}.

\section{Numerical results: Wei's method}\label{met1}

To evaluate the size ratio (\ref{gratio}) of multibranched structures as compared with linear chain, we also applied the method originally developed by Wei \cite{Wei}.
As the first step, we present the polymer structure as a mathematical graph, containing $M$ junction points (e.g. for a bottlebrush polymer with $n$ branching points of functionality $f_c$ one has $M=nf_c+2$, where two free ends of backbone are included into consideration as junction points with functionality $1$). The Kirchhoff matrix  ${\bf K}$ of the size $M\times M$ of the given graph is then defined as follows.  Its diagonal elements $K_{ii}$ equal
the degree of vertex $i$, whereas the   non-diagonal elements $K_{ij}$ equal  $-1$ when the vertices $i$ and $j$ are adjacent and $0$ otherwise.
Let $\lambda_2,\ldots,\lambda_M$ be $(M-1)$ non-zero eigenvalues of the $M\times M$
Kirchhoff matrix ($\lambda_1$ is  $0$). The
 $g$-ratio of the radii of gyration of a given branched structure and that of a linear chain with the same length is given by:
\begin{equation}
g=\frac{\sum_{j=2}^{M}1/\lambda_j^{{\rm network}}}{\sum_{j=2}^{M}1/\lambda_j^{{\rm chain}}},
\label{gwei}
\end{equation}
where $\lambda_j^{{\rm network}}$ and $\lambda_j^{{\rm chain}}$ are network and chain  
Kirchhoff matrix eigenvalues. 
	\begin{figure}[b!]\begin{center}
		\includegraphics[width=60mm]{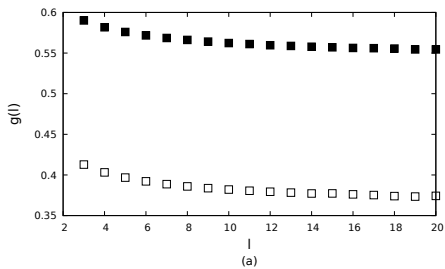}
	\end{center}
	\caption{The size ratio of the bottlebrush (filled squares) and decorated ring (open squares) at $n=3$, $f_c=2$, $f_r=0$, obtained in numerical simulations with application of Wei formula (\ref{gwei}), as functions of a single link length $l$.}
	\label{g-data}
\end{figure}

This quantity is determined within the Wei's approach by extrapolating to infinite length of links $l$, so that the total number of junction points in resulting graph (and the size of corresponding Kirchhof matrix) is $M=l((f_c+2f_r)n+n+2)$.
In our analysis, we considered  the structures with  number of  $l$ up to 20.
 The set of eigenvalues of corresponding Kirchhof matrix was evaluated for the structures presented on Fig. 1 at different $n$ and values of  $g$ were estimated  according to Eq. (\ref{gwei}).
Figure \ref{g-data} presents examples of  simulation data for the case $n=3$ and $f_c=2$, $f_r=0$. For the finite link length $l$, the values of  parameters differ from those for infinitely long polymer chains and the finite-size
deviation can be fitted by:
\begin{eqnarray}
 g(l)  =  g  + b/l,\label{fit}
\end{eqnarray}
where $a$, $b$ are constants.
The size ratio estimates were obtained by us by least-square fitting of (\ref{fit}). These are presented in Fig. \ref{g_bottle} and compared with analytical results.
Note also, that we found this method to be less reliable with increasing the number of internal closed loops in structures, due to emergence of negative eigenvalues in corresponding graphs \cite{Li01}.

\section{Results and Discussion}

\label{RD}
\begin{figure}[t!]
\begin{center}
\includegraphics[width=70mm]{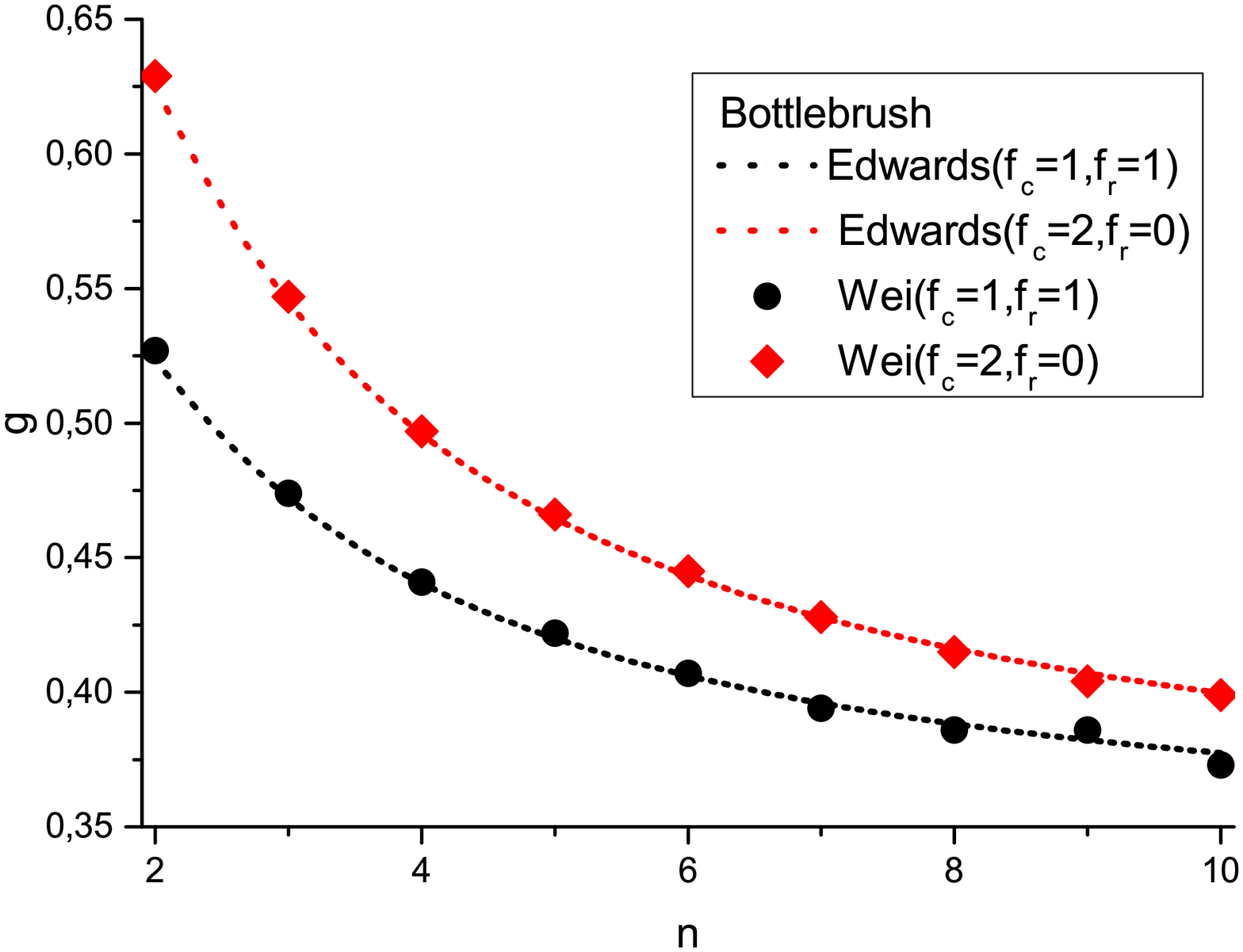}
\includegraphics[width=70mm]{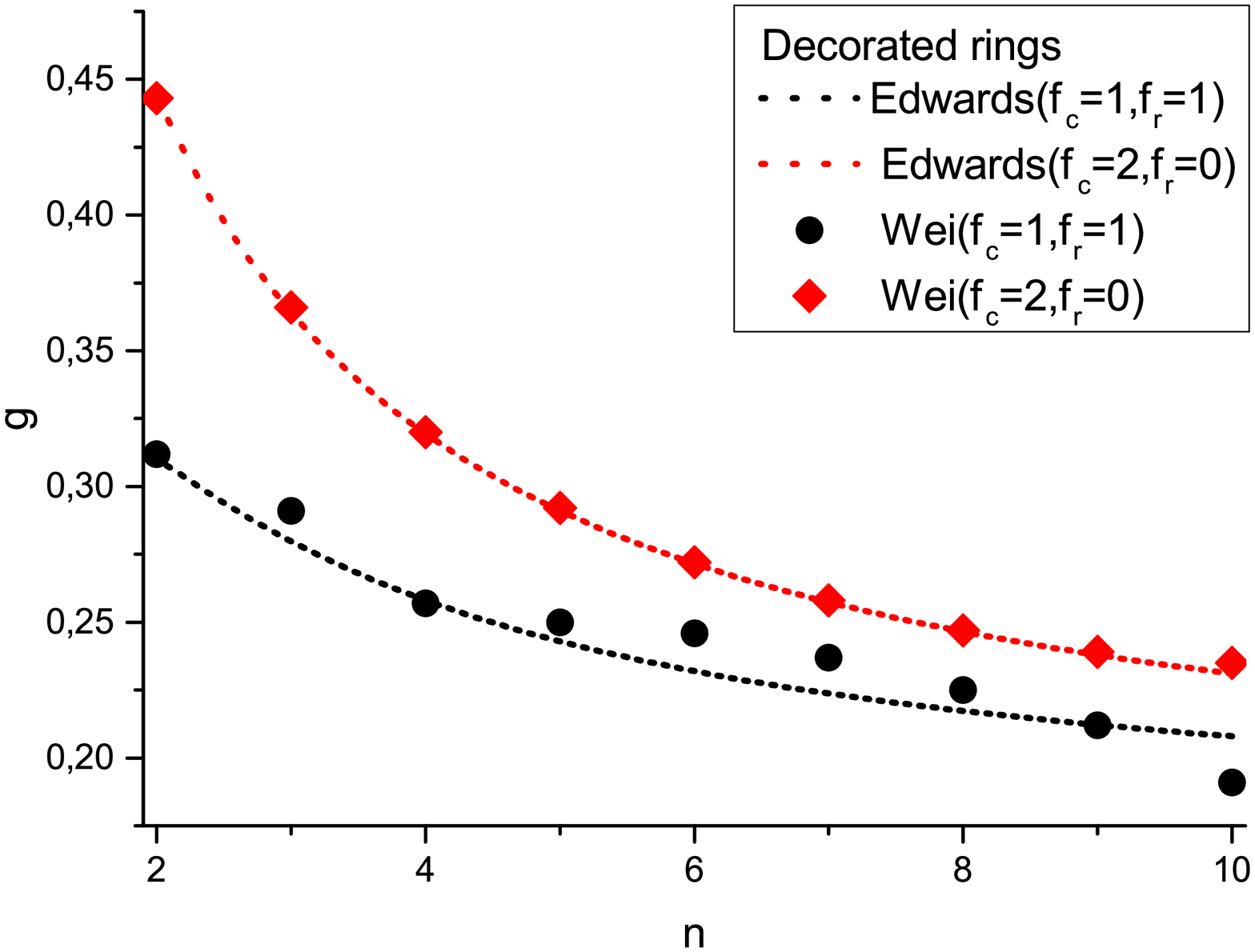}
\caption{ \label{g_bottle} The size ratio (\ref{gratio}) for bottlebrush polymers (left) and decorated rings (right) as functions of $n$ with two different sets of side chain types.}
\end{center} 
\end{figure}

\begin{figure}[b!]
\begin{center}
\includegraphics[width=70mm]{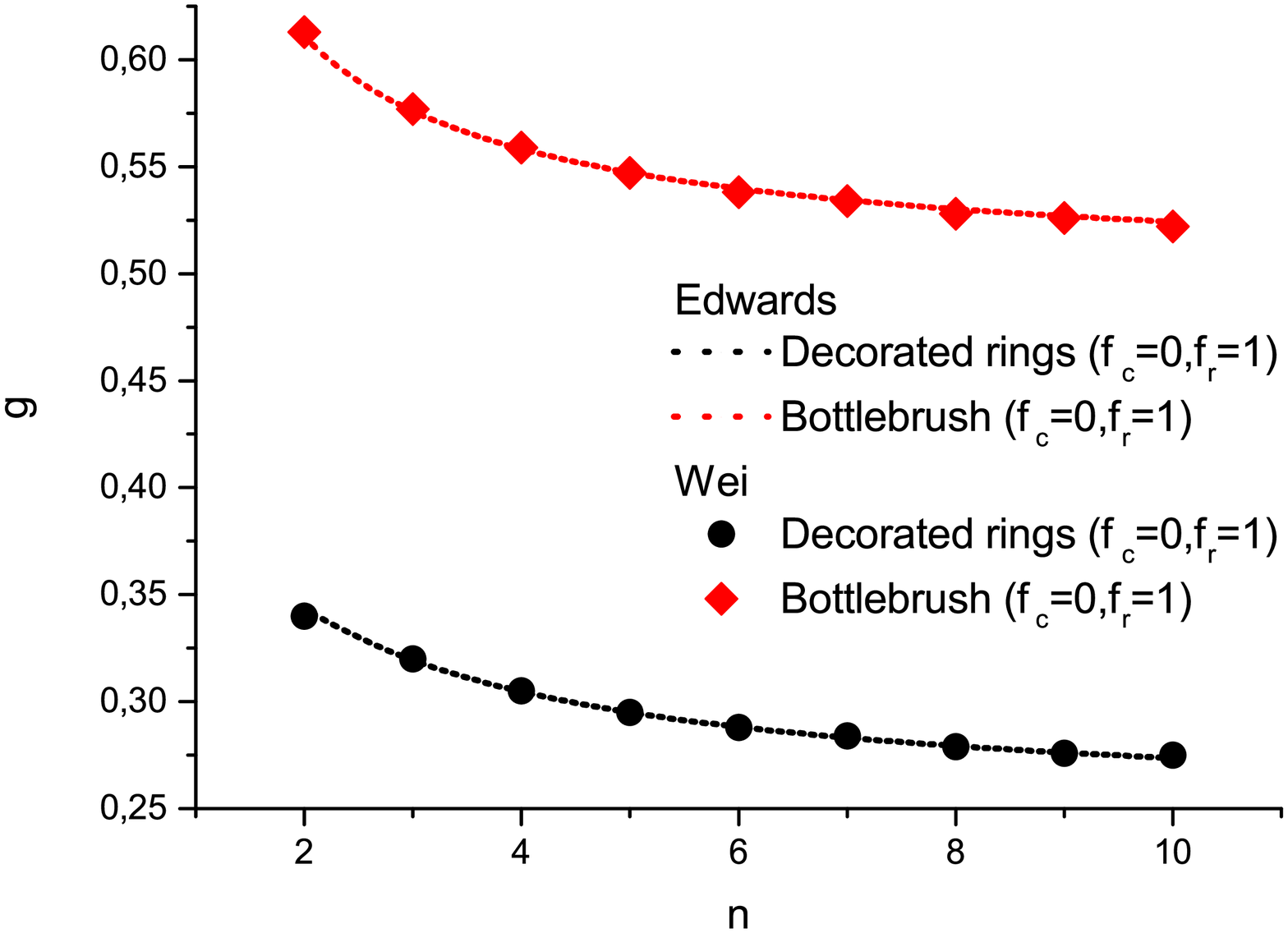}
\includegraphics[width=70mm]{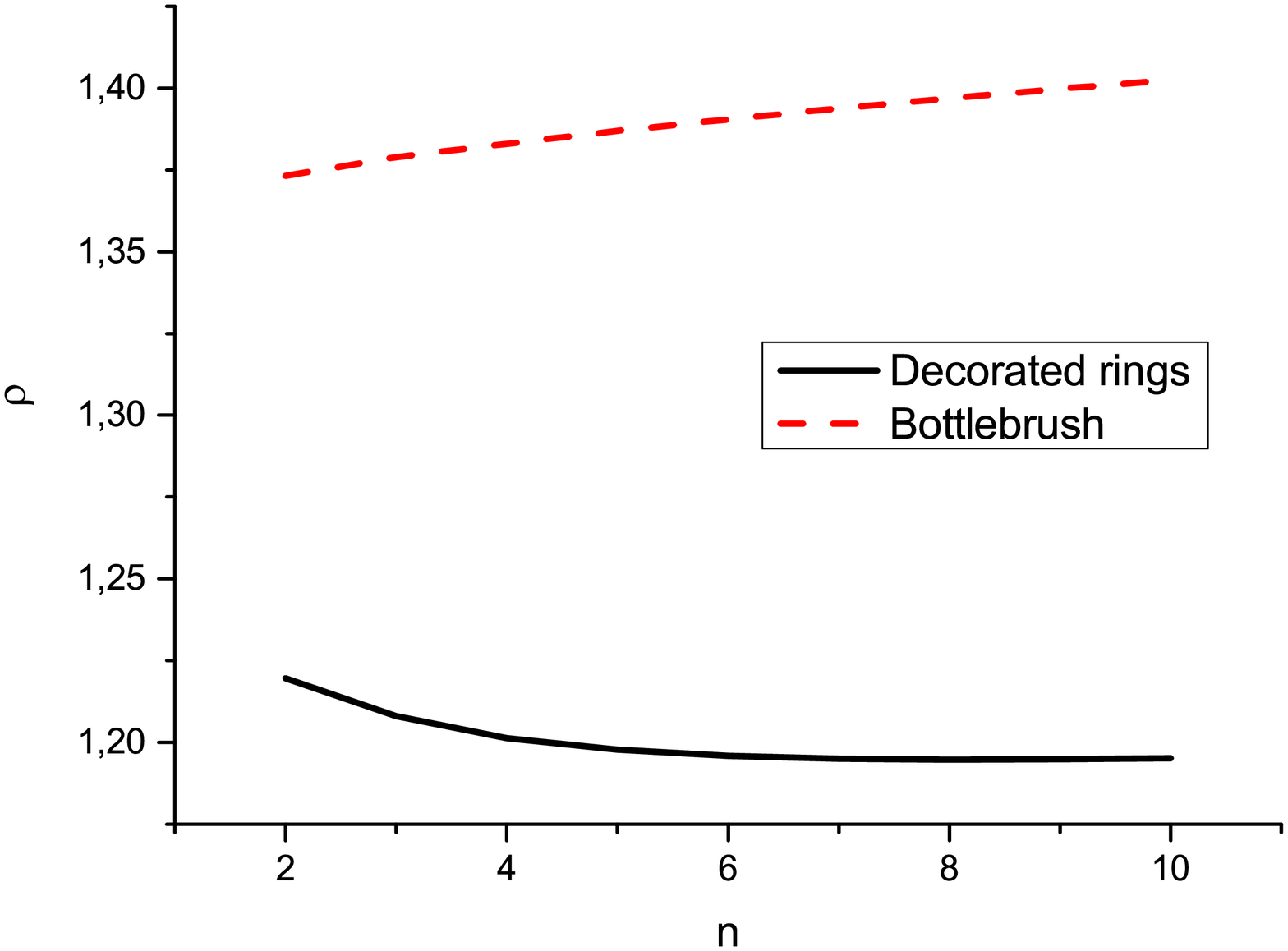}
\caption{ \label{g_ring} The size ratios (\ref{gratio}) on the left and (\ref{hratio}) on the right as functions of $n$ for toothbrushes and decorated rings with only one closed side chain.}
\end{center} 
\end{figure}
We start by considering a size ratio (\ref{gratio}) of the gyration radius of branched  polymers with $f_c$ linear branches and $f_r$ branches in form of closed loops at each of branching points, as given by (\ref{Rgbb}) and (\ref{Rgdr}) for open and closed backbones correspondingly, and that of a linear chain of the same total molecular weight. Here we limit the discussion to the case with $a=1$ that corresponds to the backbone segment between the branching points being of the same length as the side chains. Note that at $n=3$ and $f=f_c+1,f_r=0$ we restore the corresponding expression for the so-called pom-pom polymer architecture \cite{Zimm49,Radke96}, containing two branching points each of functionality $f$. As examples, the values at some fixed parameters are presented on Fig. \ref{g_bottle} in comparison with our numerical data, obtained by application of Wei's method.

At any value of $n$, the value of $g$ is smaller than $1$, which describes the shrinking of the branched structure in comparison with the linear one. This effect is increasing with increasing $n$. In the case of decorated ring (Fig. \ref{g_bottle},right) this compactification of the effective structure size at each $n$ is more pronounced. Furthermore, presence of side loops in turn leads to shrinking of the effective size of considered structures as compared with effect caused by linear side chains.  
Another important point to note is  that both analytical and numerical methods provide results that are in a good agreement, though the path integration method allows to obtained the exact values of parameters for the case of Gaussian polymers, while  the Wei's method provides only approximate values. 

{ Separately we considered the case $f_c=0$, $f_r=1$ (see Fig. \ref{g_ring}), which may be related to a problem of so-called loop extrusion (formation a set of single loops along the fibre) in chromatin  \cite{Alipour12,Fudenberg16,Goloborodko16,Mirny21}. Loop extrusion is the compaction
process, that organizes DNA.   An expected compactification (decreasing of the radii of gyration as compared with linear fiber) is observed. }

\begin{figure}[t!]
\begin{center}
\includegraphics[width=70mm]{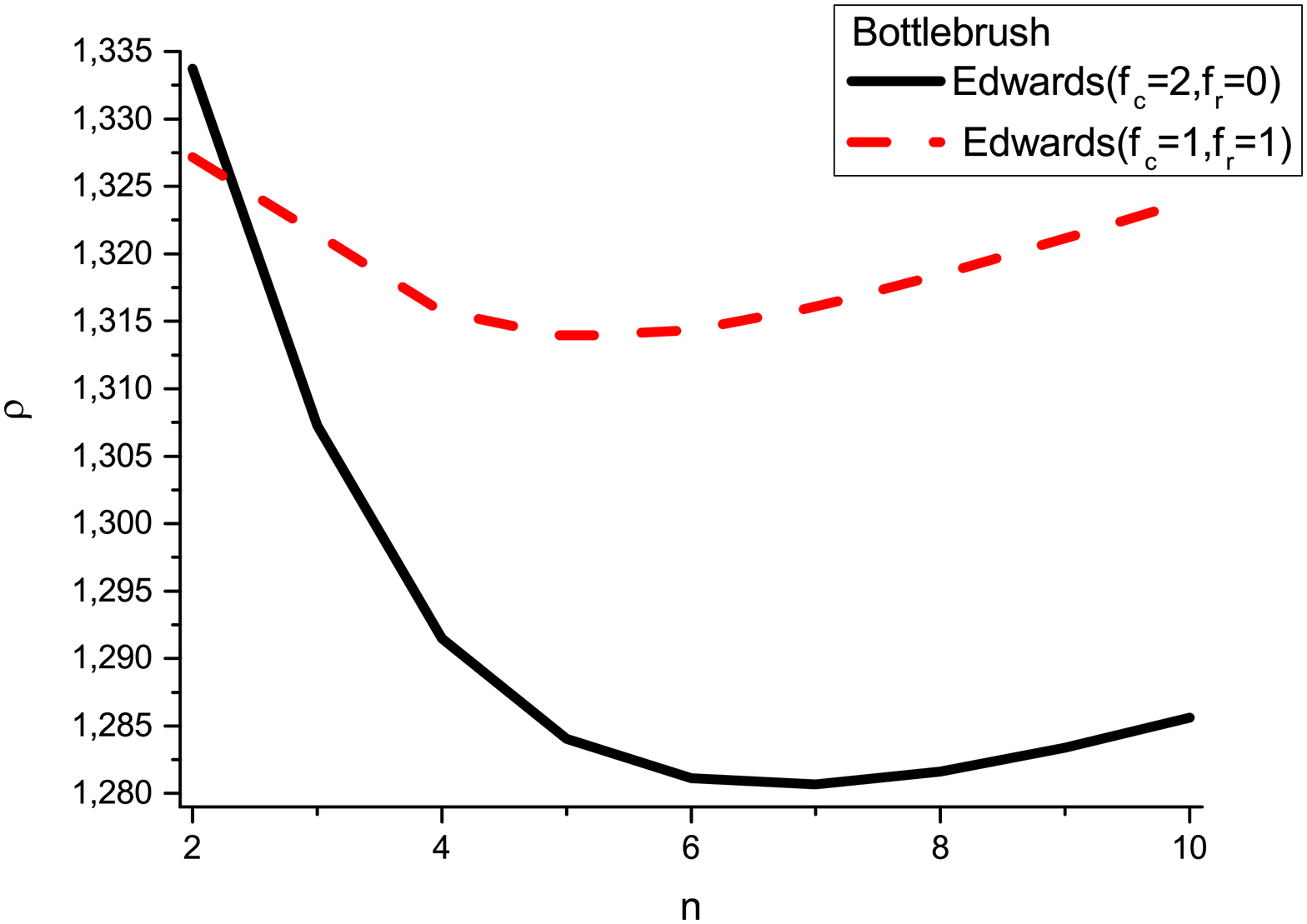}
\includegraphics[width=70mm]{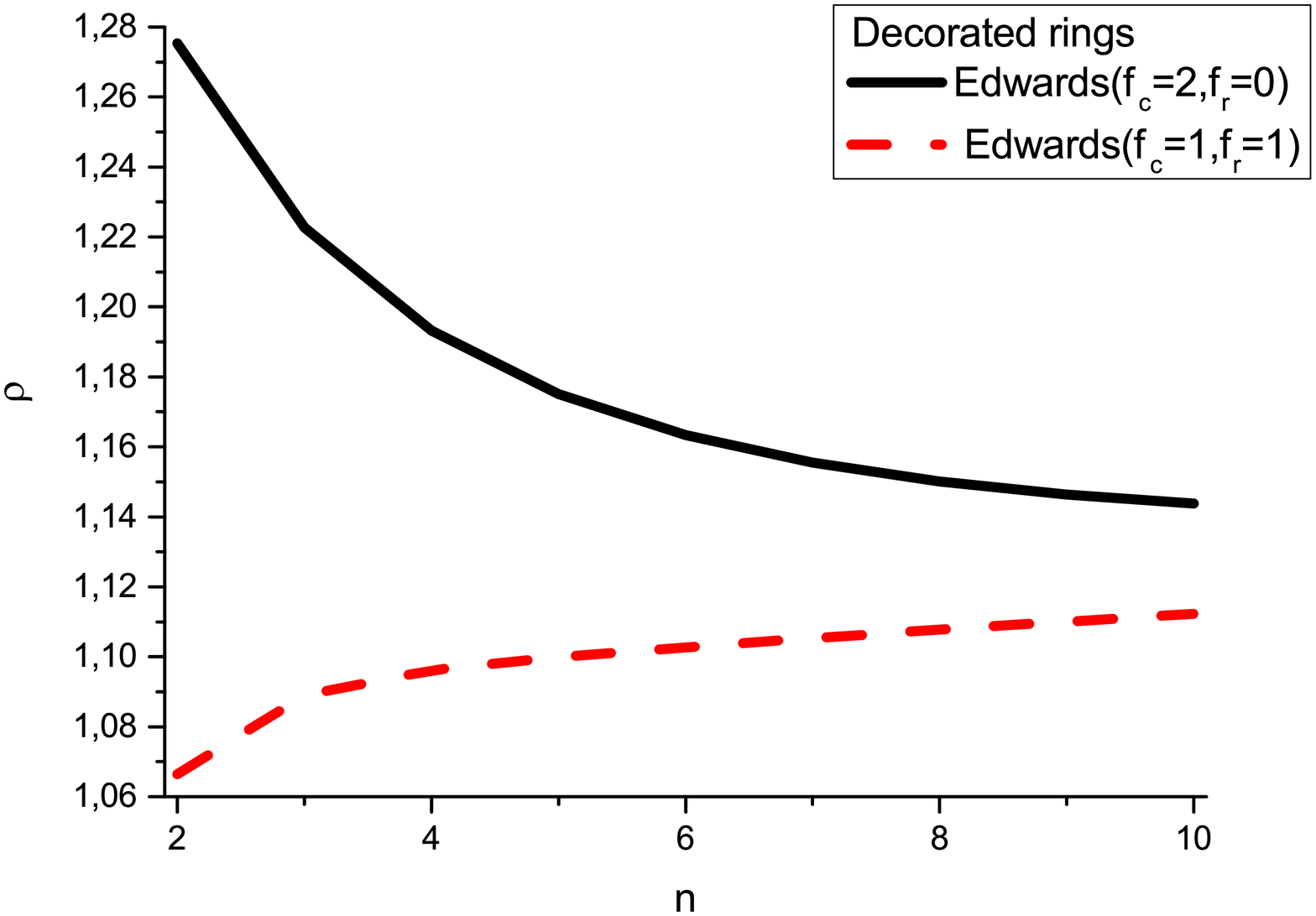}
\caption{ \label{ro_b} The size ratio (\ref{hratio})  for bottlebrush polymers (left) and decorated rings (right) as functions of $n$ with two different sets of side chain types.}\end{center}
\end{figure}

An estimate for another important size ratio, defined by (\ref{hratio}) have been evaluated on the basis of expressions (\ref{Rgbb}), (\ref{Rgdr}), (\ref{Rhbb}).
As examples, the estimates at some fixed values of parameters are presented on Fig.  \ref{ro_b}. Interesting to note, that dependence on the number of branching points $n$ is very different for bottlebrush structures with open and closed backbones. In the first case, the size ratio decreases with $n$ at small $n$  and then start to increase (left panel of \ref{ro_b}). 
%and for the large values it approaches $\rho=1.504$ %\cite{burchard,Zimm49} known for the single linear chains. 
%This behaviour may be explained by 22 stiffness at small $n$. Similarly, for the case of decorated rings (right panel of \ref{ro_b}) in the limit of large $n$ it reaches the value $\rho=1.253$ \cite{burchard}, known for simple rings. 
For the case of decorated rings, the behaviour is crucially defined by the type of side branches:  it decreases in presence of linear side chains, while looped side chains stimulate the size ratio to increase. For small values of $n$ and closed side chains, a much smaller ratio in comparison with open side chains might be explained by  more compact and thus more dense structure, so that the difference between gyration and hydrodynamic radii is very small. 
%With increase of $n$ the backbone starts to play a more %important role once again. 
%The more detail description of this behaviour will be the %subject of further research. 

\section{Conclusions}
\label{C}
The aim of the present study was to analyze the impact of the structural parameters like the number of branching points $n$ and their functionalities (number of linear $f_c$ and circular $f_r$ side branches) on the effective size measures of bottlebrush and decorated rings polymers in solutions.
In particular, the averaged size of typical conformation  of individual molecule  defines the  diffusivity of such molecules in a solvent and thus impact  their viscoelastic and phase transition behavior.

 Applying the path integration method based on Edwards model of continious chain, we evaluated exact expressions for the gyration and hydrodynamic radii of multibranched structures under considerations. The size ratios of obtained gyration radii of branched structures  (given by Eqs. (\ref{Rgbb}) and (\ref{Rgdr})) and that of the linear chain $\langle R_g^2 \rangle_{\rm chain}$ of the the same total molecular weight
were used to characterize the impact of the complex topology of structure on the effective size  in solvent (the shrinking factor). Our results quantitatively describe the shrinking of the branched structure in comparison with the linear one; this effect is increasing with increasing $n$ and is more pronounced in the case of decorated ring. Presence of side loops leads to further shrinking of the effective size of considered structures as compared with linear side chains. 
These results are confirmed also by numerical simulations performed by us within the frame of so-called Wei's method, which encounters some limitations however in presence of multiple closed loops in a structure.
Expressions for hydrodynamic radius of considered structures is evaluated as well.

The subject of the forthcoming study will be to go beyond the simplified Gaussian approximation by taking into account the excluded volume effect between monomers. This, in particular, will allow us to evaluate the impact of steric effects, caused by side branches in open and lopped form, on the effective size measure of the backbone chain and to analyze the conformational properties of considered structures more deeply.

\section*{ Acknowledgements}

Authors would like to acknowledge the support from the National
Academy of Sciences of Ukraine, Project KPKBK 6541230. 

K.H. would like to acknowledge the support from the National
Science Center, Poland (Grant No. 2018/30/E/ST3/00428).

\section*{Appendix}
The analytical expressions corresponding to all the diagrams shown on Figs. \ref{D:1}, \ref{D:2}, \ref{D:3}, needed to calculate the gyration radius of considered polymer structures, are listed below:
\begin{eqnarray}
&&\xi_{1,g}=d\frac{L^3}{6},\\
&&\xi_{2,g}=d\frac{L^3}{12},\\
&&\xi_{3,g}=d\frac{L_1L_2(L_1+L_2)}{2},\\
&&\xi_{4,g}=d\frac{L_1L_2(3L_1+L_2)}{6},\\
&&\xi_{5,g}=d\frac{L_1L_2(L_1+L_2)}{6},\\
&&\xi_{6,g}=d\left(\frac{L_1L_2(L_1+L_2)}{6}+i L_b L_1 L_2\right),\\
&&\xi_{7,g}=d\left(\frac{L_1L_2(L_1+L_2)}{2}+i L_b L_1 L_2\right),\\
&&\xi_{8,g}=d\left(\frac{L_1L_2(3L_1+L_2)}{6}+i L_b L_1 L_2\right),\\
&&\xi_{9,g}=d\left(\frac{L_1L_2(L_1+L_2)}{2}+i L_b L_1 L_2+\frac{i L_b L_1 L_2}{n}\right),\\
&&\xi_{10,g}=d\left(\frac{L_1L_2(L_1+L_2)}{6}+i L_b L_1 L_2+\frac{i L_b L_1 L_2}{n}\right),\\
&&\xi_{11,g}=d\left(\frac{L_1L_2(3L_1+L_2)}{6}+i L_b L_1 L_2+\frac{i L_b L_1 L_2}{n}\right).
\end{eqnarray}
Here, $L, L_1,\,L_2,\,L_b$ are the lengths of the segments with $L_b$ being the length between the branching points. Note that diagrams $\xi_{1,g}$ and $\xi_{2,g}$ describe contributions from ether a chain or closed ring and thus they depend on only one parameter; for diagrams $\xi_{4,g},\xi_{8,g}$ and $\xi_{11,g}$, $L_1$ is the length of the looped trajectory with a restriction point and $L_2$ of the open one  (unlike the rest of the diagrams these three are not symmetrical with respect to $L_1$ and $L_2$). The number of backbone segments  between the considered branching points is denoted by $i$, thus changing from $1$ to $n$. 

In the same way, we can present the list of expressions used to evaluate the hydrodynamic radius:
\begin{eqnarray}
&&\xi_{1,h}=C_{\xi}\frac{4L^{3/2}}{3},\\
&&\xi_{2,h}=C_{\xi}\frac{\pi L^{3/2}}{2},\\
&&\xi_{3,h}=C_{\xi}\frac{4(L_1+L_2)^{3/2}-4L_1^{3/2}-4L_2^{3/2}}{3},\\
&&\xi_{4,h}=C_{\xi}\left(L_2\sqrt{L_1}+\left(\frac{L_2^{3/2}}{2}+2\sqrt{L_2}L_1\right)\arcsin\left(\frac{L_2}{\sqrt{4L_1L_2+L_2^2}}\right)\right.\nonumber\\
&&\left.-\frac{\pi L_2^{3/2}}{4}\right),\\
&&\xi_{5,h}=C_{\xi}\pi\left(L_1\sqrt{L_2}+L_2\sqrt{L_1}-\sqrt{L_2}\sqrt{L_1}\sqrt{L_1+L_2}\right),\\
&&\xi_{6,h}=\frac{C_{\xi}}{\sqrt{4L_2L_bi+L_1L_2+L_2^2}}\left(L_2\sqrt{L_1}\right.\nonumber\\
&&\left(\arctan\left(\frac{4L_bi+L_2+\sqrt{L_2(4L_bi+L_1+L_2)}}{2\sqrt{L_1L_bi}}\right)\right.\nonumber\\
&&\left.-\arctan\left(\frac{4L_bi+L_2-\sqrt{L_2(L_bi+L_1+L_2)}}{2\sqrt{L_1L_bi}}\right)\right)(4L_bi+L_1+L_2)\nonumber\\
&&\left.-2\sqrt{L_2(4L_bi+L_1+L_2)}\left(
L_1\sqrt{L_2}\arcsin\left(\frac{L_2}{\sqrt{L_2(4L_bi+L_2)}}\right)\right.\right.\nonumber\\
&&\left.\left.+L_2\sqrt{L_1}\arctan\left(\frac{\sqrt{L_1}}{2\sqrt{L_bi}}\right)
\right)
\right),\\
&&\xi_{7,h}=C_{\xi}\frac{4}{3}\left((L_bi)^{\frac{3}{2}}-(L_bi+L_1)^{\frac{3}{2}}-(a+L_2)^{\frac{3}{2}}+(L_bi+L_1+L_2)^{\frac{3}{2}}\right),\\
&&\xi_{8,h}=C_{\xi}\frac{\sqrt{L_1}}{2}\left(\arctan\left(\frac{\sqrt{L_1}}{2\sqrt{L_bi+L_2}}\right)(4L_bi+L_1+4L_2)\right.\nonumber\\
&&\left.-\arctan\left(\frac{\sqrt{L_1}}{2\sqrt{L_bi}}\right)(4L_bi+L_1)+2\sqrt{(L_bi+L_2)L_1}-2\sqrt{L_bi L_1}\right),\\
&&\xi_{9,h}=\frac{4C_{\xi}}{3n^{\frac{3}{2}}}\left(((L_1+L_2+L_bi)n-L_bi^2)^{\frac{3}{2}}-((L_2+L_bi)n-L_bi^2)^{\frac{3}{2}}\right.\nonumber\\
&&\left.-((L_1+L_bi)n-L_bi^2)^{\frac{3}{2}}+(L_bi(n-i))^{\frac{3}{2}}\right),\\
&&\xi_{10,h}=\frac{C_{\xi}}{\sqrt{L_2n((4L_bi+L_1+L_2)n-4L_bi^2)}}\left(L_2\sqrt(L_1)\right.\times\nonumber\\
&&\left.\left(\arctan\left(\frac{4L_bi(n-i)-\sqrt{L_2}\sqrt{(4L_bi+L_1n+L_2n)-4L_bi^2}\sqrt{n}+L_2n}{2\sqrt{n(n-i)L_1iL_b}}\right)\right.\right.\nonumber\\
&&\left.-\arctan\left(\frac{(4L_bi(n-i)+L_2n+\sqrt{L_2n((4L_bin+L_1+L_2)-4L_bi^2)}}{2\sqrt{L_1iL_b(n-i)n}}\right)\right)\nonumber\\
&&\times((4L_bin+L_1+L_2)-4L_bi^2)\nonumber\\
&&+2\sqrt{L_2n((4L_bi+L_1+L_2)n-4L_bi^2)}L_1 L_2\times\nonumber\\
&&\left.\left(\frac{1}{\sqrt{L_1}}\arcsin\left(\frac{\sqrt{L_2n}}{4L_bin+L_2n-4L_bi^2}\right)\right.\right.\nonumber\\
&&\left.\left.+\frac{1}{\sqrt{L_2}}\arctan\left(\frac{\sqrt{L_1n}}{\sqrt{L_bi(n-i)}}\right)\right)\right),\\
&&\xi_{11,h}=\frac{C_{\xi}\sqrt{L_2}}{2n^{\frac{3}{2}}}\left(\sqrt{n}\arcsin\left(\frac{\sqrt{nL_2}}{\sqrt{(4L_bi+4L_1+L_2)n-4L_bi^2}}\right)\right.\nonumber\\
&&\times((4L_bi+4L_1+L_2)n-4L_bi^2)-\sqrt{n}\arcsin\left(\frac{\sqrt{nL_2}}{\sqrt{(4L_bi+L_2)n-4L_bi^2}}\right)\nonumber\\
&&\left.\times((4L_bi+L_2)n-4L_bi^2)+2\sqrt{L_2L_bin+L_1L_2n-L_2L_bi^2}n\right.\nonumber\\
&&\left.-2\sqrt(L_2L_bin-L_2L_bi^2)n\right).
\end{eqnarray}
Here, $C_{\xi}=\frac{\sqrt{2}\Gamma\left(\frac{d-1}{2}\right)}{\Gamma\left(\frac{d}{2}\right)}$.

\end{document}